\documentclass{iopconfser}

\usepackage{cite}
\usepackage{graphicx} 
\usepackage{xcolor}
\usepackage[hidelinks]{hyperref}

\usepackage[utf8]{inputenc}
\usepackage{subcaption}

\expandafter\let\csname equation*\endcsname\relax

\expandafter\let\csname endequation*\endcsname\relax 

\usepackage{amsmath}
\usepackage{amssymb}
\usepackage{hyperref}

\usepackage[normalem]{ulem}

\usepackage{fancyhdr}

\begin{document}

\title{Dynamically generated  correlations in a trapped bosonic gas via frequency quenches}

\author{Nikhil Mesquita$^{1}$, Manas Kulkarni$^{2}$, Satya N. Majumdar$^{3}$,  and Sanjib Sabhapandit$^{1}$}

\affil{$^1$Raman Research Institute, Bangalore 560080, India\\
$^2$International Centre for Theoretical Sciences, Tata Institute of Fundamental Research,\\ \hskip1.9mm Bangalore 560089, India\\
$^3$LPTMS, CNRS, Universit\'e 
Paris-Sud, Universit\'e Paris-Saclay, 91405 Orsay, France
}

\email{sanjib@rri.res.in}

\begin{abstract}
We study a system of $N$ noninteracting bosons in a harmonic trap subjected to repeated quantum quenches, where the trap frequency is switched from one value to another after a random time duration drawn from an exponential distribution. Each cycle contains two steps: (i) changing the trap frequency to enable unitary evolution under a Hamiltonian, and (ii) reapplying the original trap at stochastic times to cool the gas back to its initial state. This protocol effectively makes it an open quantum system and drives it into a unique nonequilibrium steady state (NESS). We analytically and numerically characterize the NESS, uncovering a conditionally independent and identically distributed (CIID) structure in the joint probability density function (JPDF) of the positions. The JPDF in the CIID structure is a product of Gaussians with a common random variance, which is then averaged with respect to its distribution, making the JPDF non-factorizable, giving rise to long-range emergent dynamical correlations. The average density profile of the gas shows significant deviations from the initial Gaussian shape. We further compute the order and the gap statistics, revealing universal scaling in both bulk and edge regimes. We also analyze the full counting statistics, exposing rich parameter-dependent structure. Our results demonstrate how stochastic quenches can generate nontrivial correlations in quantum many-body systems.
\end{abstract}

\section{Introduction}
\label{sec:intro}

Quantum gases have emerged as a promising ground for exploring nonequilibrium quantum dynamics \cite{bloch_rev1}, with cold atom experiments providing exceptional control over interactions \cite{Int_strength_1, Int_strength_2}, trapping geometries \cite{1D_1, 1D_2, bose_lowD}, and driving protocols \cite{quench_3, quench_1}. Among the many ways to drive such systems, quantum quenches, which are abrupt changes in a system parameter, play a central role in probing relaxation, thermalization, correlations, and phase transitions~\cite{quench_1, quench_2, quench_3, quench_4, quench_5, freq_quench_theo, freq_quench_expt}. A particularly promising class of driving protocols, which has been implemented in classical systems, involves stochastic sequences of quenches, in which the time between parameter changes is drawn from a probability distribution \cite{stochastic_quench_1, stochastic_quench_2}. Such protocols are experimentally feasible and can fundamentally alter the long-time dynamics, producing steady states that differ from both conventional thermal equilibrium and those of periodically driven systems. 
The repeated random quenches typically act as an effective driven-dissipative mechanism, rendering the system an open quantum system with an effectively non-Hermitian description. Even though the microscopic evolution between quenches is unitary, the stochastic timing of the quenches, together with averaging over realizations, leads to a breakdown of purely Hamiltonian dynamics. As a result, the system can settle into a nonequilibrium stationary state (NESS) characterized by dynamically induced correlations even among otherwise independent (noninteracting) bosons.

In this work, we investigate one such scenario: a gas of $N$ noninteracting bosons in a harmonic trap whose frequency is repeatedly switched between two values. The switching events occur after random durations sampled from an exponential distribution. Although the bosons do not interact directly, the simultaneous quench protocol induces effective all-to-all correlations. We would like to remark that the center-shift quench protocol studied in Ref.~\cite{KMS2025} is typically experimentally more demanding in cold atom setups than a frequency-switching protocol discussed in this work. This is because center-shift protocols can involve the physical transport of a lens or a magnetic coil, making it prone to excess heating due to the transfer of vibration from the transport stage \cite{Leonard_2014}.
On the other hand, controlled changes of the trap frequency can be implemented with significantly reduced heating. Trap frequencies can be altered with high precision using optical or magnetic confinement techniques. In a typical  trap, frequencies over a wide range 
(e.g., $10-500$Hz)~\cite{freq_quench_expt} can be tuned on millisecond timescales. The exponential waiting-time distributions with mean intervals from a few milliseconds to several seconds can be programmed using acousto-optic modulators, making the stochastic driving readily achievable. For ultracold bosonic gases such as $^{87}$Rb and $^{23}$Na, these parameter ranges lie well within current experimental capabilities, ensuring that the proposed frequency-quench protocol is experimentally feasible.

Remarkably, the stationary many-body joint probability density function (JPDF) exhibits a conditionally independent and identically distributed (CIID) structure, a property known from certain exactly solvable classical models with stochastic resetting~\cite{Biroli_1, Biroli_2,Biroli_3,Sanjib_2024,MSS_DD, KMS2025, stochastic_quench_1}.  This structure, where a set of random variables are independent and identically distributed, given a certain condition, makes it possible to compute a wide range of observables exactly, including density profiles, correlation functions, gap and extreme value statistics, and full counting statistics. Many of these quantities are directly measurable in ultracold atom experiments via absorption imaging~\cite{abs_img_1, abs_img_2,abs_img_3, kulkarni_2012} or quantum gas microscopy~\cite{quantum_microscopy_1, quantum_microscopy_2, quantum_microscopy_3}, suggesting that stochastic frequency quenches offer a realistic and analytically tractable route to exploring strongly correlated nonequilibrium quantum states without direct interparticle interactions. 

We briefly summarize our main findings:  (i) We find a simple quenching protocol that generates strong, all-to-all correlations in a bosonic gas, with the JPDF having a CIID structure in the steady state. (ii) We find exactly the average density profile and the statistics of the gap between successive particles. We show that the tail of the average density profile remains Gaussian, and that the tail of the gap distribution remains exponential. (iii) We show that there is a strict upper and lower bound for the number of particles that can be contained within any finite volume in the large-$N$ limit.

The paper is organized as follows. In Sec.~\ref{sec:setup}, we discuss the details of our setup and the protocol, along with details of numerical simulations employed. We obtain an interesting $N$-particle JPDF that turns out to be of CIID form. In Sec.~\ref{sec:obs}, we exploit the rich CIID structure to derive various observables. Perfect agreement is demonstrated between our analytical results and direct numerical simulations. In Sec.~\ref{sec:conc}, we summarize our results along with an outlook.

\section{Setup and Protocol}
\label{sec:setup}
Our setup consists of a gas of noninteracting bosons trapped in a harmonic potential. The protocol we employ is as follows: 
\begin{enumerate}
    \item The gas is allowed to cool to the ground state of the system, described by the Hamiltonian 
    \begin{equation}
        H(\omega_1) = \sum_{j=1}^N \left( \frac{{p}_j^2}{2m} \, + \frac{1}{2} m\, \omega_1^2 x_j^2 \right) \, ,
        \label{eq:Hamil}
    \end{equation}
    where $\omega_1$ is the trap frequency, $m$ is the mass of the particles, and $N$ are the number of particles. This ground state is described by the product state 
    \begin{equation}
\label{eq:wave_fn_in}
        \psi(x_1, x_2, \ldots, x_N) = \prod_{j=1}^N \psi_0(x_j) \ ,
        \quad \text{where}\quad 
               \psi_0(x) = \left( \frac{m \,\omega_1}{\hbar \pi} \right)^{1/4} \, \exp{\left(-\frac{m \,\omega_1 x^2}{2 \hbar} \right)} \, 
    \end{equation}
    is the ground state of the quantum harmonic oscillator.

  \item The frequency is then quenched to $\omega_2$, and the wavefunction evolves unitarily under the Hamiltonian $H(\omega_2)$, with $H(\omega)$ given in Eq.~\eqref{eq:Hamil}. This unitary evolution occurs until a stochastic time $\tau$, picked from the exponential distribution $r \, e^{-r \tau}$, where $r$ is the resetting rate. 
        
\end{enumerate} 

At the end of the duration $\tau$, the system is once again cooled to the ground state of $H(\omega_1)$. Thus, the system is reset to its initial state in Eq.~\eqref{eq:wave_fn_in}.
The entire procedure 1-2 is then repeated. 
Step~1 of our protocol is assumed to occur instantaneously. In practice, relaxation to the ground state takes a finite time, but we emulate this by disregarding the elapsed time (and making no measurements) until the system reaches the ground state with frequency $\omega_1$, as done in recent colloidal experiments~\cite{stochastic_quench_2,Faisant_2021}.
In the long-time limit, the system approaches a unique NESS, described by a CIID structure. We discuss this structure below.

\begin{figure}[t]
    \centering
\includegraphics[width=\linewidth]{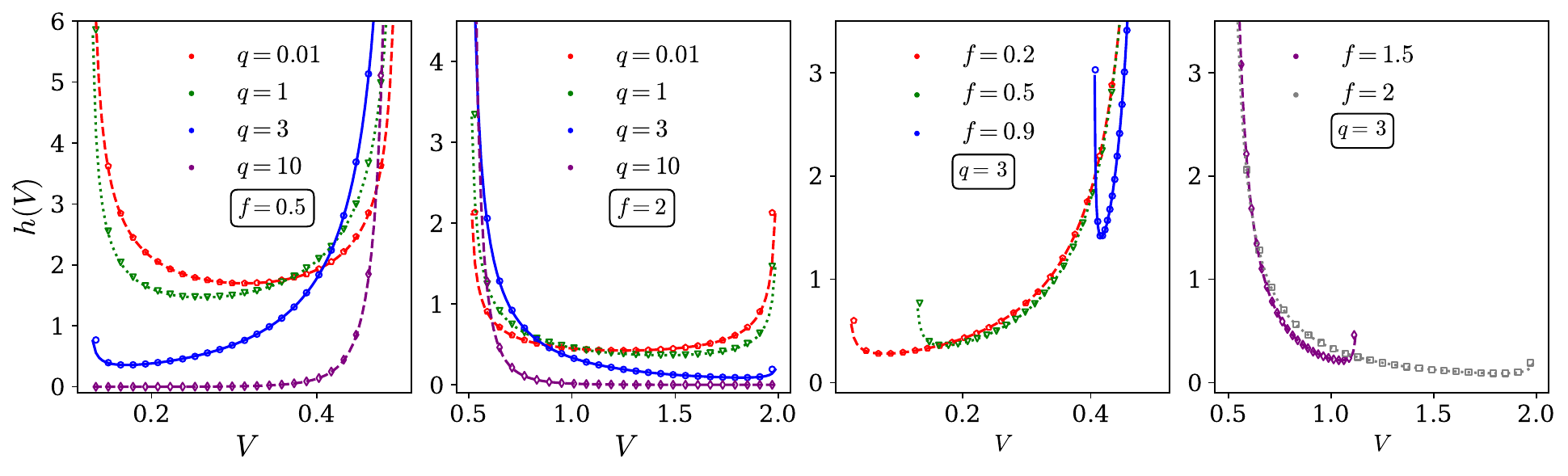}
    \caption{PDF of $V$ for various values of $f=\omega_1/\omega_2$ and $q= r/\omega_2$.
    The solid dashed lines indicate the analytical function $h(V)$ in Eq.~\eqref{eq:h_V}. The markers are obtained from direct numerical simulations  (averaged over $10^7$ realizations) described at the end of  Sec.~\ref{sec:setup}.}
    \label{fig:h_V}
\end{figure}

Under this protocol, the JPDF of the positions of the particles at any time $t$ can be written using the renewal equation~\cite{Mukherjee_quantum_reset} 
\begin{equation}
    \mathrm{Prob.}(x_1, x_2 , \ldots x_N, \, t) = e^{-r t} \prod_{j=1}^N p(x_j,t) + r \int_0^t d\tau \, e^{- r\tau} \prod_{j=1}^N \, p( x_j,\tau) \, .
    \label{eq:intro:renewal_eq}
\end{equation}
Here $\tau$ is the random time from the last resetting event, and $p(x_j, \tau) = |\psi(x_j,\tau)|^2$, with
$\psi(x,\tau) = \int_{-\infty}^\infty\, \langle x|e^{-i \tau {H} \, (\omega_2)/\hbar} |y\rangle\,  \psi_0(y)\, dy$.
We are interested in the stationary state observables, in the limit $t \rightarrow \infty$. In this limit, the first term of the renewal equation, Eq.~\eqref{eq:intro:renewal_eq} vanishes and one finds the emergence of a CIID structure,
\begin{math}
    \mathrm{Prob}.(x_1, x_2, \ldots x_N) = r \int_0^\infty d \tau \, e^{-r \tau} \, \prod_{j=1}^N \, p(x_j, \tau) \, .
\end{math}
 The strong, all-to-all correlation is clear from this unique non-factorized form of the JPDF. The explicit form of the evolved wavefunction
$\psi(x,\tau)$ can be obtained by using the quantum propagator, 
\begin{equation}
    K(x,y ,\tau) \equiv \langle x| e^{-i \tau {H} \, (\omega_2)/\hbar} |y\rangle= \frac{m\, \omega_2}{2 i \pi \hbar \, \sin(\omega_2 \tau)} \ \exp{\left( \frac{i\, m \,\omega_2}{2 \hbar \sin(\omega_2 \tau)} \left[ (x^2 + y^2)\cos(\omega_2 \tau) - 2 xy\right]\right)} \, .
    \label{eq::kernel_shm}
\end{equation}
Since both $K(x,y,\tau)$ and $\psi_0(y)$ are Gaussian,  $\psi(x,\tau)=\int_{-\infty}^\infty  K(x,y,\tau)\, \,  \psi_0(y) \, dy$ remains Gaussian at all times. Consequently,  the single particle conditional PDF $p(x,\tau)=|\psi(x,\tau)|^2$ is again a  Gaussian, 
\begin{equation}
    p(x,\tau)\equiv p(x|V(\tau)) = \frac{1}{\sqrt{2 \pi \sigma_1^2 V(\tau)}} \, \exp{\left(- \frac{x^2}{2 \sigma_1^2 V(\tau)} \right)} \, ,
\quad \text{where} \quad \sigma_1 = \sqrt{\frac{\hbar}{m \,\omega_1}}    ~~,
\label{eq:p(x|V)}
\end{equation}
and $\tau$ appears through the variable $V(\tau)$ as 
\begin{equation}
    V(\tau) =\frac{1}{4} \left[(1+f^2) + (1-f^2) \cos(2\omega_2 \tau )\right]\, , \ \text{with  $f = \frac{\omega_1}{\omega_2}$. }
    \label{eq:def_V}
\end{equation}
It is evident from Eq.~\eqref{eq:def_V} that 
$V\in (V_1,V_2)$ with 
\begin{equation}
    V_1 =  \frac{1}{2}\min[ f^2,1], \ \text{and } \,  
    V_2 =  \frac{1}{2}\max[f^2,1] \, .
    \label{eq:V1V2}
\end{equation}
Therefore, NESS JPDF $\mathrm{Prob}.(x_1, x_2, \ldots x_N)$ can be expressed in terms of the random variance $V$ as 
\begin{equation}
    \mathrm{Prob}.(x_1, x_2 ,\ldots x_N) = \int_{V_1}^{V_2} dV \, h(V) \, \prod_{j=1}^N p(x_j|V) \,,\quad\text{where}\quad  h(V) = r \int_0^\infty d\tau \, e^{-r \tau} \, \delta[V- V(\tau)] \, 
    \label{eq:CIID_NESS}
\end{equation}
is the distribution of $ V$, and $p(x|V)$ is  given in Eq.~\eqref{eq:p(x|V)}. To proceed further, we must obtain $h(V)$ in Eq.~\eqref{eq:CIID_NESS}, 
with $V(\tau)$ given in Eq.~\eqref{eq:def_V} and $\tau$ drawn from the exponential distribution $r\, e^{-r \tau}$. The evaluation of $h(V)$ requires us to express $\tau$ as a function of $V$ from Eq.~\eqref{eq:def_V}. To do this, we must find the roots of 
\begin{equation}
    \cos(2 \, \omega_2 \tau) = \alpha(V) = \frac{4 \, V - f^2 - 1}{1 - f^2} \, ,
    \label{eq:alp_def}
\end{equation}
for $\tau>0$.
The smallest root, $\tau_1$ is given by, $\tau_1 = \cos^{-1}(\alpha)/(2\, \omega_2)$.
Since $\cos(2 \omega_2 \tau)$ has a period of $\pi/\omega_2$ there are roots to Eq.~\eqref{eq:alp_def} at 
\begin{math}
  \tau_{2s+1} = {s \pi}/{\omega_2} \, + \tau_1,\,\,  \text{with}\,\, s=1,2, \dotsc
\end{math}
There is an additional root, denoted by $\tau_{2s}$, within each period. The two roots $\tau_{2s-1}$ and $\tau_{2s}$ within each period are symmetrically placed with respect to $(2s-1)\pi/(2 \omega_2)$, where $s=1, 2,\dotsc$.  This yields
\begin{math}
    \tau_{2s} = {s \pi}/{\omega_2} \, - \tau_1,\,\,\text{with}\,\, s=1,2, \dotsc \, 
\end{math}
Summing over all the roots, the PDF $h(V)$ in Eq.~\eqref{eq:CIID_NESS} becomes
\begin{equation}
    h(V) = \frac{ q}{2\sinh(q \pi/2)} \, \frac{\cosh\left[\frac{q}{2} \cos^{-1}(-\alpha) \right]}{\sqrt{(V_2-V)(V-V_1)}} \quad \text{with}\quad \alpha(V) = \frac{4 V -1 -f^2}{1-f^2} \, , 
    \label{eq:h_V}
\end{equation}
and $q = r/\omega_2$, $f=\omega_1/\omega_2$, and $V_1$ and $V_2$ are given in Eq.~\eqref{eq:V1V2}. Clearly, $h(V)$ has square-root divergences at both supports. 
In Fig.~\ref{fig:h_V}, we plot the PDF $h(V)$ in Eq.~\eqref{eq:h_V} for different values of $f$ and $q$ and compare with direct numerical simulations. We find excellent agreement between the two.

Henceforth, we set $\sigma_1 =1$ and compute various macroscopic and microscopic observables for this model. These observables have been validated through numerics obtained by the method described below.\\

\noindent
\textit{Numerical method employed.--- }
The numerical data is obtained as follows: (i) We first pick a random number $\tau$ from an exponential distribution  $r e^{-r \tau}$. (ii)  Next we obtain $V(\tau)$ defined in Eq.~\eqref{eq:def_V}. To find the distribution of $V(\tau)$ in Fig~\ref{fig:h_V}, $10^7$ realizations of $V$ are obtained, and the data are binned and averaged.
For the observables: (a) We pick $\tau$ from the exponential distribution and find the corresponding $V(\tau)$. (b)  For each $V$, we generate $N$ Gaussian random numbers, taking $V$ as the variance. This procedure (a)-(b) is repeated for  $10^5-10^6$ number of realizations, and the observables are computed, binned, and averaged over all the realizations.
 
\section{Observables}
\label{sec:obs}
\subsection{Average Density Profile} 
\begin{figure}[t]
    \centering
\includegraphics[width=\linewidth]{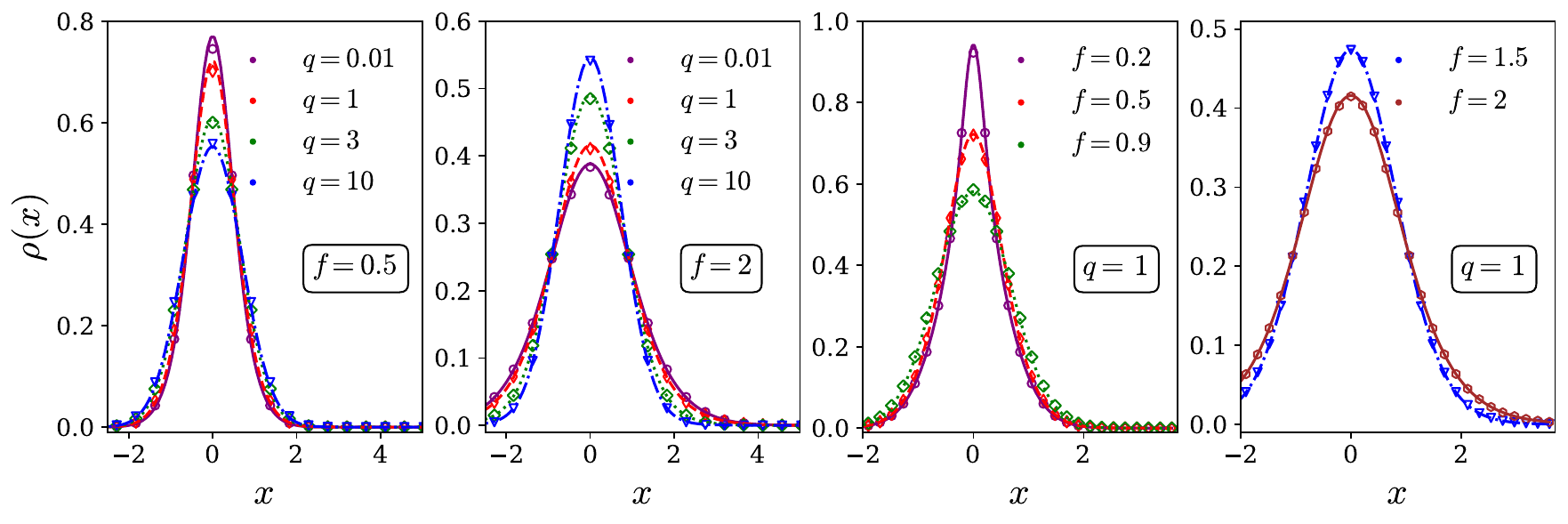}
\includegraphics[width=0.6\linewidth]{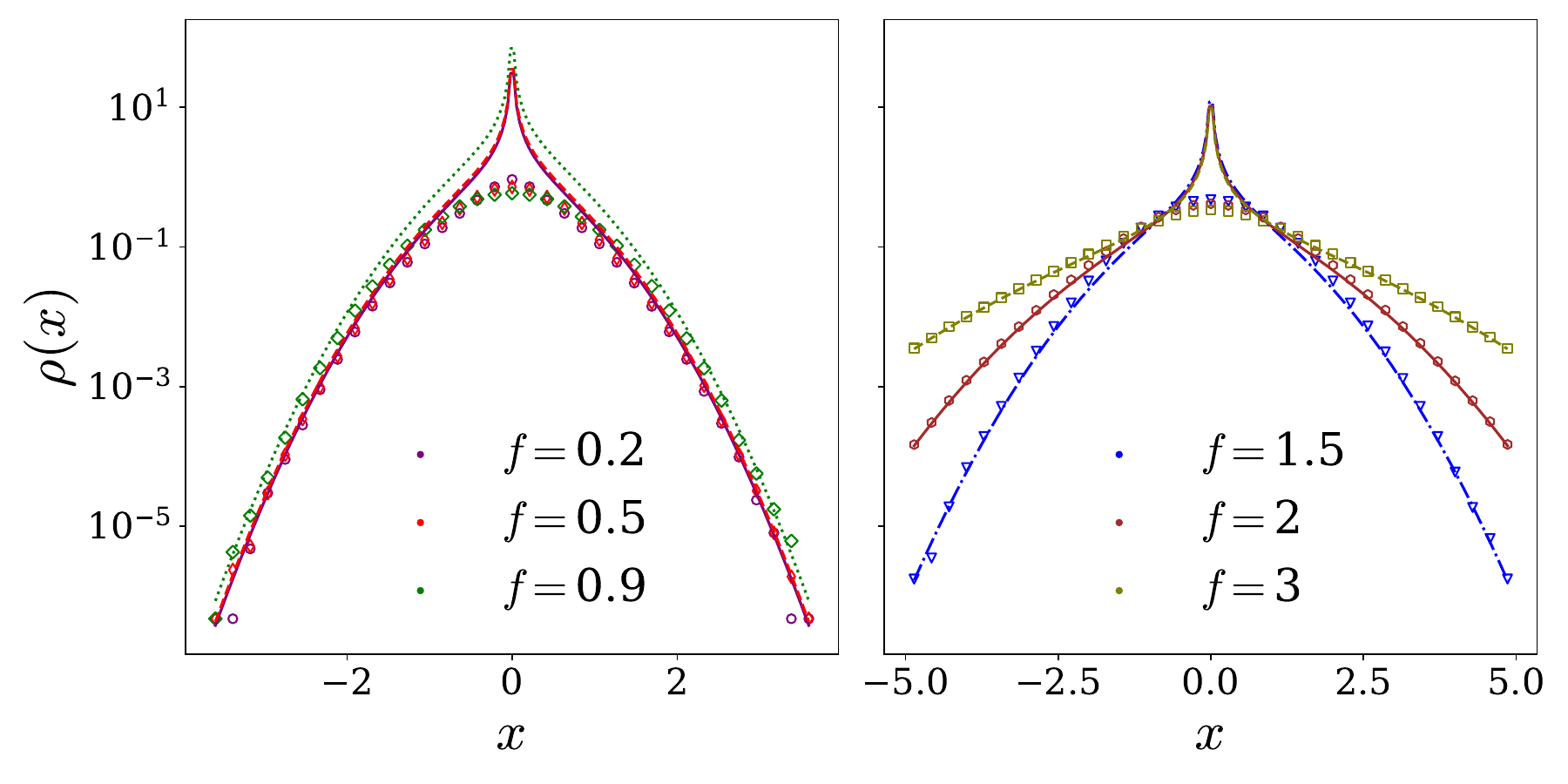}
    \caption{The top panel plots of the average density profile for various $f=\omega_1/\omega_2$ and $q=r/\omega_2$, with the solid lines indicating the analytical function in Eq.~\eqref{eq:den}. The markers are obtained numerically by the procedure described in Sec.~\ref{sec:setup}, with $\sigma_1 =1$, and averaged over $10^7$ realizations. The bottom panel highlights 
    the asymptotic tail of $\rho(x)$. The central cusp-like feature in analytics is an artifact of the approximate expression in  Eq.~\eqref{eq:den_tail}, which is valid only far from the center (large $x$).
    }
    \label{fig:den}
\end{figure}

The average density profile $\rho(x) = N^{-1} \langle \sum_{i=1}^N \delta(x_i-x) \rangle$ is nothing but the marginal single particle distribution. From the CIID structure in Eq.~\eqref{eq:CIID_NESS}, it follows that 
\begin{equation}
    \rho(x) = \int_{V_1}^{V_{2}} dV \, h(V) \, \frac{e^{-x^2/(2V)}}{\sqrt{2 \pi V}} \, .
    \label{eq:den}
\end{equation}
Although $h(V)$ has a finite support, the average density profile in Eq.~\eqref{eq:den} is supported over the entire real line.
The comparison of numerical data and the analytical average density profile (with the integral evaluated numerically)  in Fig.~\ref{fig:den} shows an excellent agreement.

The asymptotic behavior of $\rho(x)$  can be obtained by noting that due to the essential singularity of $V$ in $e^{-x^2/(2V)}$, for $x \gg \sqrt{V_2}$, the most dominant contribution to the integral in Eq.~\eqref{eq:den} comes from near the upper limit $V_2$. Therefore, we set $V=V_2-\epsilon$ and keep only the leading order terms in $\epsilon$ in the integrand in Eq.~\eqref{eq:den}. From Eq.~\eqref{eq:h_V}, for $\epsilon\to 0$, we get  
\begin{equation}
h(V_2-\epsilon)\simeq \frac{A_f}{\sqrt{\epsilon}} \quad \text{where} \quad A_f=\frac{q \cosh\big[(q/2)\, \cos^{-1}(-\alpha(V_2))\big]}{ \sqrt{2|1-f^2|}\,\sinh(q\pi /2)} .
\label{eq:Af}
\end{equation}
Hence, for $x \gg \sqrt{V_2}$, Eq.~\eqref{eq:den} yields
\begin{equation}
    \rho(x) \sim A_f\, \frac{e^{-x^2/(2V_2)}}{\sqrt{2 \pi V_2}} \, \int_0^\infty \frac{d\epsilon}{\sqrt{\epsilon}}\, e^{-\epsilon \, x^2/(2 V_2^2)}  = A_f\, \frac{\sqrt{V_2}}{|x|}\, e^{-x^2/(2V_2)}\, ,
    \label{eq:den_tail}
\end{equation}
where we have 
extended the upper limit of the integral to $\infty$ (since the integrand decays very fast).
Hence, the tail remains Gaussian, albeit with a different variance and a logarithmic correction.
The form of the tail in Eq.~\eqref{eq:den_tail} is validated by simulation in Fig.~\ref{fig:den}.

\subsection{Order Statistics}
\begin{figure}[t]
    \centering
    \includegraphics[width=\textwidth]{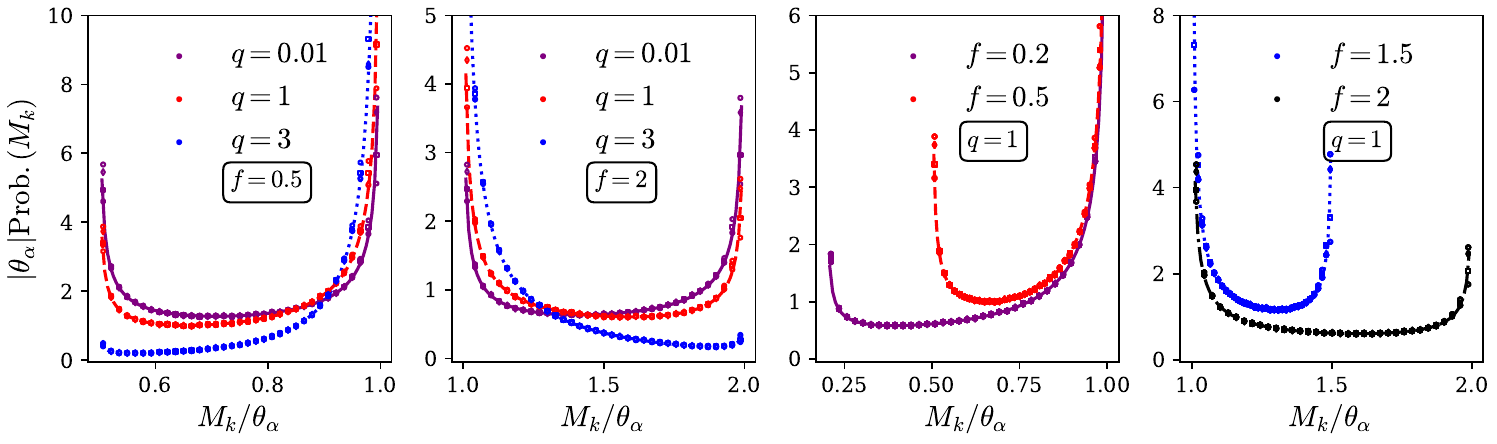} 
    \caption{Plot of the scaled collapsed distribution of $M_k/\theta_\alpha$ for $\alpha =$ 0.0001 (square), \,  0.1 (circle), \, 0.3 (diamond), \, 0.45 (hexagon),  and various values of parameters $f=\omega_1/\omega_2$ and $q=r/\omega_2$. The markers indicate the numerical data (details in Sec.~\ref{sec:setup}) for parameters $\sigma_1=1$, $N=10^6$, averaged over $10^6$ realizations. The solid dashed lines are the analytical plots of the scaling function $z \, h(z^2/2) \equiv |\theta_\alpha|  \, \mathrm{Prob}.(M_k)$ [Eq.~\eqref{eq:os}].}
    \label{fig:Os}
\end{figure}

The order statistics $\mathrm{Prob}.(M_k = w)$ represent the position of the $k$-th furthest particle along the line, where $k=1$ corresponds to the rightmost particle, $k=2$ to the second rightmost particle, and so on.
The order statistics for this class of models are obtained by first computing these for a fixed $V$. Since the conditional single particle distributions are Gaussian, and IID, an asymptotic analysis leads to the conditional distribution being $\mathrm{Prob}.(M_k|V) = \delta(M_k - \sqrt{2 V} \mathrm{erfc}^{-1}(2 \alpha) )$, with $\alpha = k/N$~\cite{MSS_DD, Biroli_1}. Thus, averaging this over $V$ leads to 
\begin{equation}
    \mathrm{Prob}.(M_k = w) = \frac{w}{\theta_\alpha^2} \, h{\left( \frac{w^2}{2 \, \theta_\alpha^2} \right)} \, ,
    \label{eq:os}
\end{equation}
with $h(z)$ defined in Eq.~\eqref{eq:h_V}, and $\theta_\alpha =\mathrm{erfc}^{-1}(2 \alpha)$. This scaling function is valid for both the bulk as well as the edge, and has finite support $M_k/\theta_\alpha \in \left( \min(f,1), \, \max(f,1) \right)/\sqrt{2}$. Note that for $\alpha >1/2$, $\theta_\alpha$ is negative, and the corresponding $M_k$'s have support over negative values, but are still described by the scaling function Eq.~\eqref{eq:os}. Moreover, we unravel the interesting feature that although the density profile is supported over the real line, the order statistics has finite support~\cite{Sanjib_2024,KMS2025}.
These results are validated in Fig.~\ref{fig:Os}.

\subsection{Gap Statistics} 
\begin{figure}[t]
    \centering  \includegraphics[width=\textwidth]{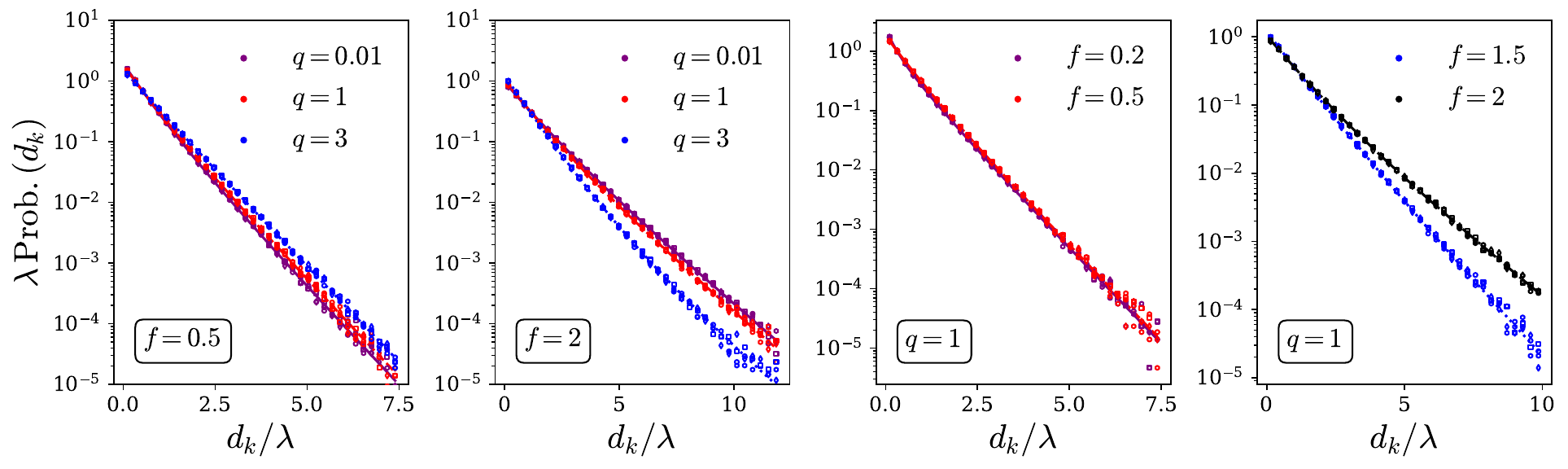}
    \caption{Plot of the scaled collapsed distribution of $d_k/\lambda$ for $\alpha =$ 0.0001 (square), \,  0.1 (circle), \, 0.3 (diamond), \, 0.45 (hexagon) and various values of parameters $f = \omega_1/\omega_2$ and $q = r/ \omega_2$. The markers indicates numerically obtained data (see Sec.~\ref{sec:setup}),  for $\sigma_1=1$, $N=10^6$, averaged over $10^6$ realizations. The solid dashed lines represent the scaling function $F(z)$ given in Eq.~\eqref{eq:gap_scalfn} with integral evaluated numerically.}
    \label{fig:gap}
\end{figure}

We compute the statistics of the gap between successive particles, $d_k = M_k - M_{k+1}$. This is done by fixing $V$, and analyzing the IID random variables~\cite{Biroli_1}. After averaging over $h(V)$, this leads to 
\begin{equation}
    \mathrm{Prob}. (M_k - M_{k+1}=d_k) = \frac{1}{\lambda} \, F{\left( \frac{d_k}{\lambda}\right)} \, , \quad     F(z) = \int_{V_{1}}^{V_{2}} dV \,    h(V) \frac{e^{-z/\sqrt{V}}}{\sqrt{V}} \, .
    \label{eq:gap_scalfn}
\end{equation}
with $\lambda = N^{-1} \sqrt{2 \pi } \, e^{\theta_\alpha^2}$, $\theta_\alpha = \mathrm{erfc}^{-1}(2 \alpha)$.
The gap statistics results are validated in Fig.~\ref{fig:gap}. The integrand in the scaling function $F(z)$ in Eq.~\eqref{eq:gap_scalfn} has a similar essential singularity as that in the average density profile in Eq.~\eqref{eq:den}. Therefore, for $z \gg \sqrt{V_2}$, a similar analysis as in Sec.~\ref{sec:obs}, yields
\begin{equation}
    F(z)  \sim \frac{A_f}{\sqrt{V_2}}\, e^{-z/\sqrt{V}}\, \int_0^\infty 
\frac{d\epsilon}{\sqrt{\epsilon}}\, e^{-\epsilon \, z/(2 V_2^{3/2})} =  \sqrt{2 \pi}A_f\,  
    \frac{V^{1/4}}{\sqrt{z}}  \, e^{-z/\sqrt{V}}\, ,
\end{equation}
where $A_f$ is given in Eq.~\eqref{eq:Af}.
Thus, the tail of the scaling function of the gap statistics remains exponential, with the quenching protocol introducing a subleading logarithmic correction.

\begin{figure}[t]
    \centering
\includegraphics[width=0.65\linewidth]{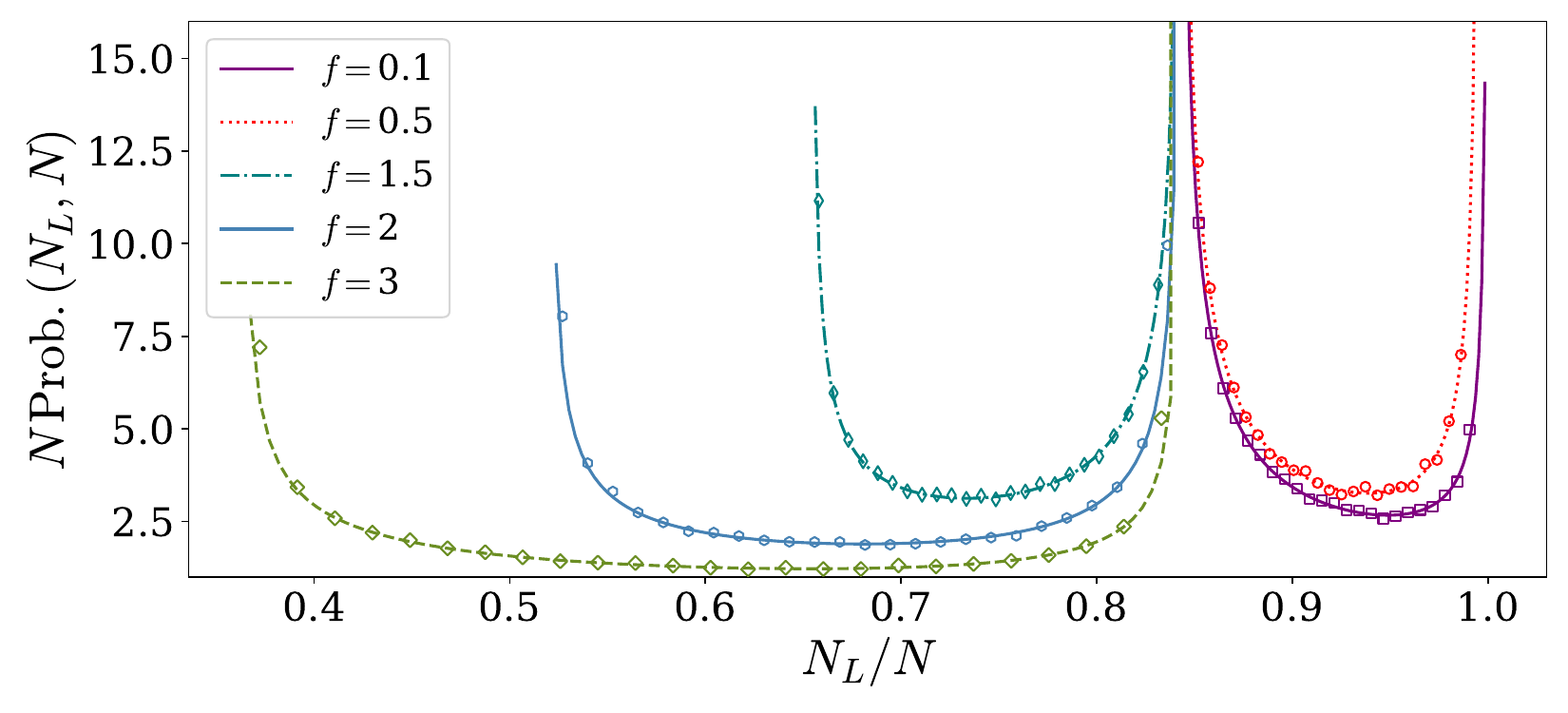}
    \caption{FCS plot for $L=1$, $q= r/\omega_2=1$ and various $f = \omega_1/\omega_2$. The markers indicate simulations (details in section~\ref{sec:setup}) for $\sigma_1=1$, $N=10^5$, averaged over $10^5$ realizations. The solid dashed curves are analytical plots of the function $\mathrm{Prob}.(\kappa,L)$ in Eq.~\eqref{eq:fcs}.}
    \label{fig:FCS}
\end{figure}

\subsection{Full Counting Statistics} 
Exploiting the CIID structure, given a $V$, the probability of having $N_L$ particles within $[-L,L]$ is a binomial distribution, $\mathrm{Prob.}(N_L,N|V) = \binom{N}{N_L} \, \phi(L|V)^{N_L} [1-\phi(L|V)]^{N-N_L}$. Here $\phi(L|V) = \int_{-L}^L dx \, p(x|V)$, where $p(x|V)$ is defined in Eq.~\eqref{eq:p(x|V)}. In the $N \rightarrow \infty$ limit, we know that the binomial distribution converges to a Gaussian distribution, and consequently the fraction of particles contained between $[-L,L]$ is a Dirac delta function about $\phi(L|V) = \mathrm{erf}(L/\sqrt{2V})$~\cite{MSS_DD}. Averaging this over $V$ gives us the required FCS, which reads  
\begin{equation}
    \mathrm{Prob}.(\kappa,L) = \frac{\sqrt{\pi} L^2}{2 \, [\mathrm{erf}^{-1}(\kappa)]^3} \, \exp{\left( [\mathrm{erf}^{-1}(\kappa)]^2\right)} \, h{\left( \frac{L^2}{2 [\mathrm{erf}^{-1}(\kappa)]^2}\right)} \, .
    \label{eq:fcs}
\end{equation}
This is supported over $[\kappa_{\min}, \kappa_{\max}]$, with $\kappa_{\max} = \max[ \mathrm{erf(L)}, \, \mathrm{erf}(L f^{-1})]$ and   $ \kappa_{\min} =     \min[ \mathrm{erf(L)}, \, \mathrm{erf}(L f^{-1})]$.
Hence, in any volume, no more than $\kappa_{\max}$ and no less than $\kappa_{\min}$ fraction of particles can be contained in the thermodynamic limit of $N\to\infty$. These findings are validated in Fig.~\ref{fig:FCS}.

\section{Conclusion and Outlook}
\label{sec:conc}
We have studied a system of $N$ noninteracting bosons in a harmonic trap subjected to stochastic frequency quenches, in which the trap frequency alternates between two chosen values after random exponentially distributed durations. Despite the absence of direct interactions, the protocol induces strong dynamical correlations, driving the system to a unique NESS whose JPDF displays a CIID structure. The system and the driving protocol yield an effectively non-Hermitian setup. This stands out as a rare example of an open quantum system that is both analytically tractable and implementable in cold-atom experiments. The CIID mathematical structure we obtained enables exact computation of several observables, including density profiles, gap and order statistics, and full counting statistics, revealing both universal and parameter-dependent features.

Some future questions include incorporating inherent interactions, either perturbatively or within mean-field frameworks, to probe the consequences of departure from a CIID structure~\cite{Biroli_dysonBM}. The intricate interplay between inherent interactions and dynamically generated correlations is expected to yield highly non-trivial steady states.
It will also be interesting to explore non-exponential waiting-time distributions~\cite{Evans_2020}, which could lead to qualitatively different steady states. Finally, given that our setup and protocol in particularly suitable for ultracold atom experiments,  combining state-of-the-art absorption imaging techniques~\cite{abs_img_1, abs_img_2,abs_img_3, kulkarni_2012} also with high-resolution detection methods, such as quantum gas microscopy~\cite{quantum_microscopy_1, quantum_microscopy_2, quantum_microscopy_3}, could enable direct observation of the predicted correlations and statistics, paving the way towards controlled engineering of correlated open quantum systems.

\section{Acknowledgements}
MK acknowledges the support of the Department of Atomic Energy, Government of India, under project no. RTI4001. SNM and SS acknowledge the support from the Science and Engineering Research Board (SERB, Government of India), under the VAJRA faculty scheme (No. VJR/2017/000110). SNM acknowledges
support from ANR Grant No. ANR-23-CE30-0020-01
EDIPS.

\providecommand{\newblock}{}

\end{document}